\documentclass[manuscript, nonacm]{acmart}
\usepackage[utf8]{inputenc}
\usepackage{multirow}
\usepackage{graphicx}
\usepackage{longtable}

\title{AI in Computational Thinking Education in Higher Education: A Systematic Literature Review\footnotemark[1]}
\author{Ebrahim Rahimi}
\affiliation{
    \institution{Open University, Computer Science department}
    \country{The Netherlands}
}
\email{ebrahim.rahimi@ou.nl}

\author{Clara Maathuis}
\affiliation{
    \institution{Open University, Computer Science department}
    \country{The Netherlands}
}
\email{clara.maathuis@ou.nl}
\date{Sep 2025}

\begin{document}
\begin{abstract}
\textbf{Abstract:} Computational Thinking (CT) is a key skill set for students in higher education to thrive and adapt to an increasingly technology-driven future and workplace. While research on CT education has gained remarkable momentum in K-12 over the past decade, it has remained under-explored in higher education, leaving higher education teachers with an insufficient overview, knowledge, and support regarding CT education. The proliferation and adoption of artificial intelligence (AI) by educational institutions have demonstrated promising potential to support instructional activities across many disciplines, including CT education. However, a comprehensive overview outlining the various aspects of integrating AI in CT education in higher education is lacking.

To mitigate this gap, we conducted this systematic literature review study. The focus of our study is to identify initiatives applying AI in CT education within higher education and to explore various educational aspects of these initiatives, including the benefits and challenges of AI in CT education, instructional strategies employed, CT components covered, and AI techniques and models utilized.

This study provides practical and scientific contributions to the CT education community, including an inventory of AI-based initiatives for CT education useful to educators, an overview of various aspects of integrating AI into CT education—such as its benefits and challenges (e.g., AI's potential to reshape CT education versus its potential to diminish students' creativity)—and insights into new and expanded perspectives on CT in light of AI (e.g., the 'decoding' approach alongside the 'coding' approach to CT).

\end{abstract}
\maketitle
\footnotetext[1]{A poster based on this paper was accepted and published in the Proceedings of the 30th ACM Conference on Innovation and Technology in Computer Science Education (ITiCSE 2025), DOI: https://doi.org/10.1145/3724389.3730775.}







\section{Introduction}

Computational Thinking (CT) is an important thinking skill set required by students to thrive and adapt to the future \cite{hsu2018learn}. CT comprises various concepts and practices, including abstraction, data collection, analysis, and representation, algorithms and procedures, problem decomposition, automation, parallelization, and simulation \cite{barr2011bringing}. CT is defined as a means of 'solving problems, designing systems, and understanding human behavior by drawing on the concepts fundamental to computer science' \cite{wing2006computational} and an essential thinking skill and analytical ability for the digital age, alongside reading, writing, and arithmetic, for everyone, not just computer scientists \cite{wing2006computational}. Although the integration of CT into the K-12 curriculum has gained significant attention worldwide, it remains largely underexplored in higher education \cite{lyon2020computational}. 



Artificial Intelligence (AI) has appeared as a potential game changer in (CT) education \cite{chen2020artificial, lin2021pdl,tedre2021ct}. AI could benefit various aspects of teaching and learning processes, improve the efficiency and effectiveness of education administration, support customized and personalized learning, and develop meaningful learning experiences and materials \cite{timms2016letting}. In the context of CT education in K-12, several AI-driven initiatives have been launched to integrate AI into CT education, including Wolfram Programming Lab, IBM Watson, Google’s Teachable Machine, and Snap! \cite{tedre2021ct}. Furthermore, AI's profound influence on computation and CT education has prompted new conceptualizations of CT, such as CT 2.0 \cite{tedre2021ct}.  



Teachers, as the main change agents in education, play an indispensable role in integrating AI into CT education to help their students acquire and deepen CT competencies. However, the educational and technological specifications, benefits, and challenges of integrating AI into CT education remain largely unclear and underexplored, particularly in the context of higher education \cite{hsu2018learn}. This raises a dire need for higher education instructors to seek, explore, and evaluate various AI-based initiatives for teaching CT, and to equip themselves with effective AI-based instructional knowledge and strategies to enhance their teaching. 

To start recognizing and addressing this need, we conducted a systematic literature review (illustrated in Fig. \ref{fig:literaturestrategy} and detailed in the next section) aiming to explore various aspects of existing applications of AI in CT education within higher education. This study identified and reported the benefits and challenges of integrating AI into CT education, examined instructional strategies, addressed CT components (in terms of CT concepts, practices, and perspectives \cite{brennan2012new}), and utilized AI methods and techniques. It is noteworthy that since the beginning of 2023, the AI discourse in research and practice across various domains, including education and CT education, has been dominated by Generative AI (e.g., ChatGPT) and Large Language Models. Since research and practice on the application of Generative AI and LLMs in education are still in their infancy and require more time to establish consistency and reliable evidence, we opted to exclude them from this study, as explained in the following section.

\section{Literature Review strategy}
\begin{figure}[ht]
  \centering
  \includegraphics[width=\linewidth]{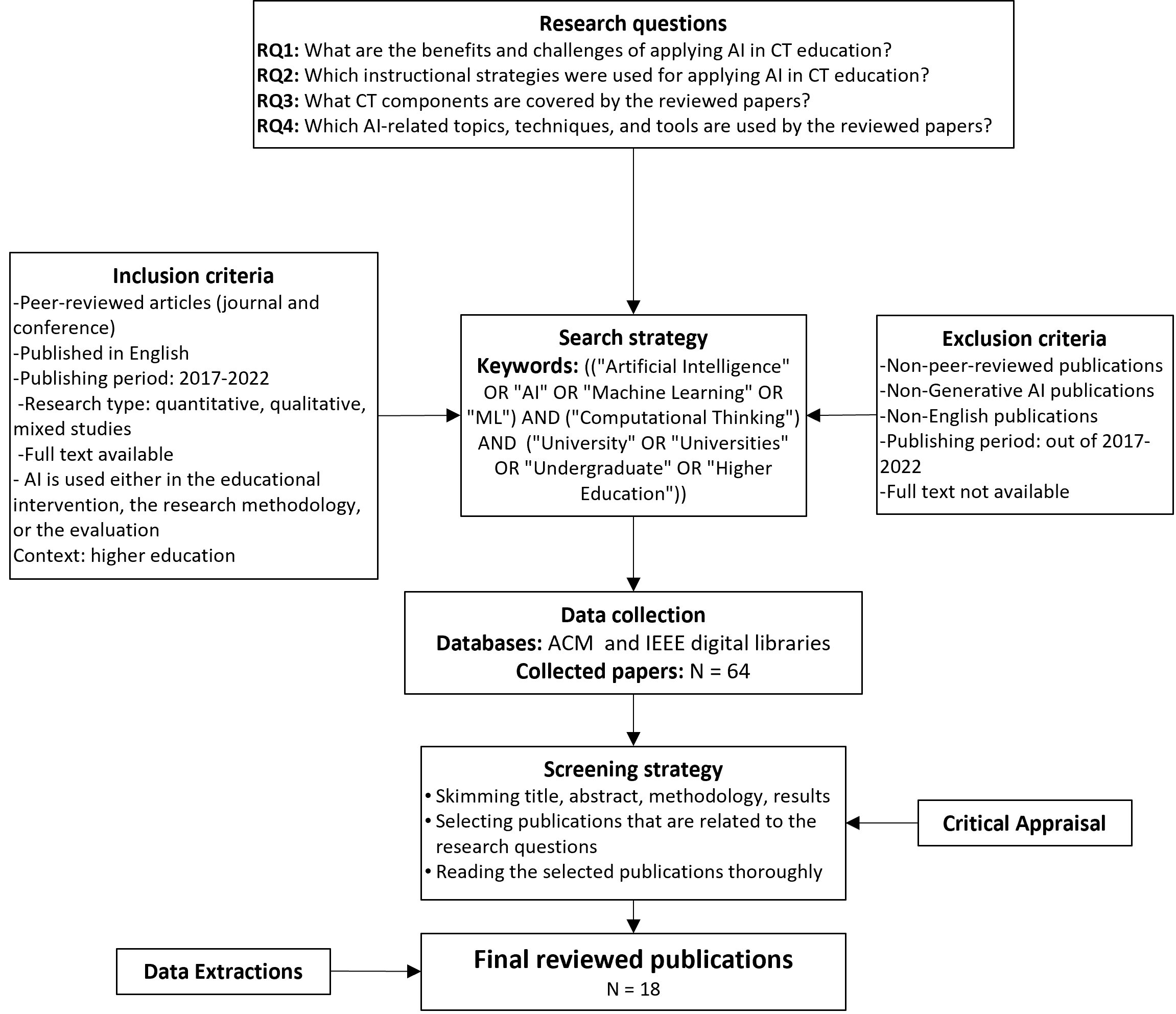}
  \caption{The followed systematic literature review process}
  \label{fig:literaturestrategy}
\end{figure}

We followed the systematic literature review strategy suggested by Lockwood and Oh \cite{lockwood2017systematic} consisting of seven steps as follows:
(1) developing structured questions, (2) defining inclusion and exclusion criteria, (3) developing a search strategy, (4) critical appraisal, (5) data extraction, (6) analysis of extracted data, and (7) presentation of the findings. Figure \ref{fig:literaturestrategy} depicts an overview of the literature review process we followed.  

\subsection*{Inclusion and Exclusion criteria}
We included peer-reviewed journal articles and conference publications, published in English between 2017 and 2022, that focused on the context of higher education and for which the full text was available. We deliberately selected the timeframe from 2017 to 2022 to exclude publications related to Generative AI and Large Language Models, which began to proliferate significantly across various venues starting in 2023. The exclusion of Generative AI and LLM papers from this study is due to several reasons, including the immaturity of evidence regarding their usefulness for education and the lack of methodological consistency in how these technologies can and should be applied in educational contexts. The included publications may have employed qualitative, quantitative, or mixed-methods research strategies, incorporating AI in either their educational design and intervention, the research methodology, or evaluation in the context of CT education.

\subsection*{Search Strategy and data collection}

To formulate the search strategy, the initial step involved identifying keywords and pertinent synonyms related to AI and CT in higher education. To address the AI aspect, we utilized the keywords 'Artificial Intelligence', 'AI', 'Machine Learning', and 'ML' employing the boolean operator "OR". To address the Computational Thinking part, we included 'computational thinking' keyword. To address the higher education context, we utilized the keywords 'University,' 'Universities,' 'Undergraduate,' and 'higher education'. We chose to use the ACM and IEEE digital libraries for our search due to their comprehensive coverage and high recognition in the AI and Computational Thinking disciplines. 

\subsection*{Critical appraisal}
Running the above search resulted in a list of N = 64 publications for the critical appraisal, conducted by two researchers of this study, collaboratively. Researcher 1, with an educational and research background in software engineering and programming education, and researcher 2, with an educational and research background in AI and programming education. First, all selected publications were downloaded and stored in a shared online repository accessible to both researchers. Next, the title, abstract, authors, venue, and publication year of each article were extracted and stored in an Excel spreadsheet. Subsequently, both reviewers read the title and abstract of each publication, adding notes and making the decision on inclusion or exclusion based on four criteria: (1) alignment with the search strategy's inclusion requirements, (2) relevance to the formulated research questions, (3) use of AI in methodology or evaluation, and (4) focus on computational thinking or programming as the core subjects. When more information was needed to decide on a specific publication, the researchers studied the full text. Finally, both researchers shared and discussed their notes and decisions. Facing diverse opinions regarding the selection of a publication, the researchers debated and clearly articulated their reasoning to reach a consensus. This evaluation process resulted in a definitive list of 18 publications selected for this review (N = 18), presented in Table \ref{tab:reviewedpapers}.


\begin{table}
 \caption{List of the reviewed papers}
    \label{tab:reviewedpapers}
 \fontsize{8pt}{8pt}\selectfont
\begin{tabular}{|p{2.0in}|p{3.0in}|}
 \hline
 \textbf{Authors; Year; Reference} & \textbf{Title} \\ 
 \hline
 Han, Kexin and Zhang, Lingzhi; 2021; \cite{han2021exploration} & Exploration on the Path of Cultivating Innovative Talents under the Background of Intelligent Era \\
 \hline
 Liu, Song and Xie, Xiaoyao; 2021; \cite{liu2021ai} & AI Quality Cultivation and Application Ability Training for Normal University Students\\ 
 \hline

 Ng, Andrew Keong and Atmosukarto, Indriyati et al.; 2021; \cite{ng2021development} & Development and implementation of an online adaptive gamification platform for learning computational thinking \\ 
 \hline
 Prisco, Andr{\'e} and dos Santos, Rafael et al.; 2017; \cite{prisco2017using} & Using information technology for personalizing the computer science teaching\\ 
 \hline
 Miljanovic, Michael A and Bradbury, Jeremy S; 2020; \cite{miljanovic2020gidgetml} & GidgetML: an adaptive serious game for enhancing first year programming labs\\ 
 \hline
 Sung, Isaac and Berland, Matthew; 2017; \cite{sung2017forest} & Forest friends demo: A game-exhibit to promote computer science concepts in informal spaces\\ 
 \hline
 Bart, Austin Cory and Kafura, Dennis et al.; 2018;\cite{bart2018reconciling} & Reconciling the promise and pragmatics of enhancing computing pedagogy with data science \\ 
 \hline
 Katchapakirin, Kantinee and Anutariya, Chutiporn; 2018; \cite{katchapakirin2018architectural} & An architectural design of scratchthai: A conversational agent for computational thinking development using scratch\\ 
 \hline
 Hemberg, Erik and Kelly, Jonathan and O'Reilly, Una-May; 2019; \cite{hemberg2019domain} & On domain knowledge and novelty to improve program synthesis performance with grammatical evolution\\ 
 \hline
 Shih, Wen-Chung; 2019; \cite{shih2019integrating} & Integrating computational thinking into the process of learning artificial intelligence\\ 
 \hline
 Figueiredo, Jos{\'e} and Garc{\'\i}a-Pe{\~n}alvo, Francisco Jos{\'e}; 2019; \cite{figueiredo2019teaching} & Teaching and learning strategies of programming for university courses\\ 
 \hline
 Figueiredo, Jos{\'e} and Lopes, Noel and Garc{\'\i}a-Pe{\~n}alvo, Francisco Jos{\'e}; 2019; \cite{figueiredo2019predicting} & Predicting student failure in an introductory programming course with multiple back-propagation\\ 
 \hline
 Brown, Onis and Roberts-Elliott et al.; 2020; \cite{brown2020abstract} & Abstract visual programming of social robots for novice users\\
 \hline
 Zimmermann-Niefield, Abigail and Polson, Shawn et al.; 2020; \cite{zimmermann2020youth} & Youth making machine learning models for gesture-controlled interactive media\\ 
 \hline
 Emerson, Andrew and Geden, Michael et al.; 2020; \cite{emerson2020predictive} & Predictive student modeling in block-based programming environments with Bayesian hierarchical models\\ 
 \hline
 Lall{\'e}, S{\'e}bastien and Yal{\c{c}}{\i}n, {\"O}zge Nilay and Conati, Cristina; 2021; \cite{lalle2021combining} & Combining Data-Driven Models and Expert Knowledge for Personalized Support to Foster Computational Thinking Skills\\ 
 \hline
 Lin, Shu and Meng, Na; 2021; \cite{lin2021pdl} & PDL: scaffolding problem solving in programming courses\\
 \hline
 Lunding, Mille Skovhus and Gr{\o}nb{\ae}k, Jens Emil Sloth et al.; 2022; \cite{lunding2022exposar} & ExposAR: Bringing Augmented Reality to the Computational Thinking Agenda through a Collaborative Authoring Tool\\ 
 \hline
 
 \end{tabular}

\end{table}

\subsection*{The analysis process}
We followed an open thematic qualitative content analysis \cite{elo2008qualitative} approach to answering the research questions as follows: we studied the full text of the selected papers and iteratively identified and grouped the reported benefits and challenges posed by AI to CT education concerning the first question, next to the utilized instructional strategies to apply AI in CT education to answer research question 2. Addressing research question 3 involved two steps. First, given the diversity of CT concepts and definitions proposed in the literature, we adopted the CT components defined by \cite{brennan2012new}, which consist of CT concepts, practices, and perspectives as the overall categories. Then, we used these categories to analyze and group the CT components addressed in the reviewed papers. To answer research question 4, we employed a bottom-up inductive approach without relying on any pre-defined categories of AI techniques.


\section{The benefits and challenges (RQ1)}
Table \ref{tab:benefits} depicts the identified benefits and challenges of applying AI to CT education. AI has been advocated for its \textit{high potential for reshaping education} in general, and CT in particular. Han and Zhang \cite{han2021exploration} highlighted that the unique intelligent and automatic identification, collection, processing, and screening of information about students and their learning next to the adaptive perception capabilities of AI can provide unprecedented opportunities to cultivate innovative talents within higher education. In this regard, AI has been praised for its strong potential to transform educational processes within higher education, fostering more autonomous lifelong learning through cooperation between machine and human \cite{liu2021ai}.

\begin{table}[]
    \caption{Identified benefits and challenges of applying AI in CT education}
    \label{tab:benefits}
    \fontsize{8pt}{8pt}\selectfont
    \begin{tabular}{|p{3.5in}|p{1.0in}|}
     \hline
     \textbf{Identified benefits} & \textbf{Papers} \\
     \hline
      High potential to reshape CT education   & \cite{han2021exploration,liu2021ai} \\
      \hline
       Personalizing teaching and learning of CT & \cite{prisco2017using,katchapakirin2018architectural,figueiredo2019teaching,figueiredo2019predicting,emerson2020predictive}\\
      \hline
     Addressing issues of students' diverse CT background& \cite{ng2021development,miljanovic2020gidgetml}\\
          \hline
    Building students' CT knowledge and skills& \cite{katchapakirin2018architectural,shih2019integrating,brown2020abstract,zimmermann2020youth,lunding2022exposar}\\
          \hline
      Motivating and promoting the acquisition and practice of CT skills& \cite{sung2017forest,bart2018reconciling,katchapakirin2018architectural,shih2019integrating,brown2020abstract,lunding2022exposar,zimmermann2020youth}\\
       \hline
\multicolumn{2}{c}{} \\
\hline
\textbf{Identified challenges} & \textbf{Papers} \\
     \hline
      Diminishing Students' Creativity and Innovation Ability   & \cite{han2021exploration} \\
      \hline
     The complexity of characterizing students' CT background and competencies& \cite{figueiredo2019teaching,lalle2021combining}\\
          \hline
    Knowledge and expertise requirements for AI development& \cite{brown2020abstract,bart2018reconciling}\\
      \hline

    \end{tabular}
    
\end{table}

\textit{Personalizing teaching and learning of CT} using AI is a frequently reported benefit of AI for CT education by the reviewed papers. Several papers emphasize the benefits of using AI techniques and applications for CT education, such as recommendation systems, student profiling, and modeling using neural network predictive models and Bayesian hierarchical modeling. These techniques can significantly facilitate personalized teaching and learning in CT and programming. In particular, these AI techniques can benefit CT education by understanding each student's profile, respecting their individual characteristics \cite{prisco2017using}, addressing unique learning needs, interests, and difficulties \cite{katchapakirin2018architectural}, predicting potential student struggles in advance, and implementing adaptive activities to support students' learning \cite{figueiredo2019teaching,figueiredo2019predicting,emerson2020predictive}.

AI can help to \textit{address issues related to students' diverse CT background and knowledge}. 
Students attending CT and programming education may come from diverse
academic backgrounds with heterogeneous prior CT knowledge and experience. It has been reported that AI-based courses such as CTQ (i.e., Computational Thinking Quest, a novel online adaptive gamified course \cite{ng2021development}) and adaptive games such as adaptive GidgetML \cite{miljanovic2020gidgetml} can serve as a bridge to narrow the CT heterogeneity gap within the group of students.

The benefits of AI for \textit{building students' CT knowledge and skills} have been highlighted by several papers ( \cite{katchapakirin2018architectural,shih2019integrating,brown2020abstract,zimmermann2020youth,lunding2022exposar}). In \cite{brown2020abstract}, the authors presented
a high-level visual programming system which enabled novices to
design robot tasks by incorporating social behavioral cues to improve the perception of robots as social agents in a public museum. Zimmermann-Niefield et al. \cite{zimmermann2020youth} combined a beginner-level machine learning (ML) modeling toolkit with a programming tool to investigate how learners create and remix projects to incorporate custom ML-based gestural inputs. As reported by this study, combining programming and ML allowed the participants to build creative projects controlled by personalized
input, including modalities like gesture, speech, and live video, and by integrating ML models of their own gestures into programming projects. Lunding et al. in \cite{lunding2022exposar} attempted to extend the CT domain by bringing AR (Augmented Reality) to the CT agenda by developing and evaluating a collaborative authoring tool called ExposAR. According to this study, the ExposAR-based collaborative authoring process was beneficial in supporting students in understanding AR concepts and challenging their perspectives on AR.    

 The influence of AI-based initiatives on \textit{motivating and promoting} students to learn CT concepts were highlighted by several papers. This influence includes promoting and facilitating learning of CT Concepts in informal and public spaces such as museums (e.g., Forest Friends strategic video game \cite{sung2017forest}), via solving contextualized real-world data-driven problems (e.g., \cite{bart2018reconciling}), supporting collaborative learning and extending CT learning beyond schools' walls \cite{katchapakirin2018architectural,shih2019integrating,brown2020abstract,lunding2022exposar}, and providing creative learning experiences \cite{zimmermann2020youth}. 

Regarding the challenges of applying AI to CT education, Han and Zhang \cite{han2021exploration} warned about \textit{students' over-reliance on AI}, which could diminish their creativity and innovative abilities. This might happen due to the replacement of students' innovative thinking with advanced neural networks and other AI techniques with powerful and complicated information processing and problem-solving capabilities, which can simulate the human brain's operation mechanisms.


The \textit{complexity of characterizing students' CT background and competencies} is reported as a challenge by some papers \cite{figueiredo2019teaching,lalle2021combining}. Most AI-based initiatives in CT and programming education create and use predictive student models based on their CT and programming background and performance for personalizing and adapting CT teaching and learning. However, most often these predictive student models are based on simplifying or even invalid assumptions, such as assuming a normal response distribution and homogeneous student characteristics. These simplified assumptions can limit or negatively impact the predictive performance and accuracy of these models and the personalized and adaptive learning tools built upon them \cite{lalle2021combining}.

\textit{Knowledge and expertise requirements} of applying AI in CT education is another highlighted challenge. Developing AI-based initiatives (e.g., making social robots and sustaining their use in public spaces such as museums as a means to extend and motivate CT education in public \cite{brown2020abstract}) requires not only AI expertise but also the domain expertise of the staff of the target Environment. To cope with the limited access to AI experts after launching such initiatives, they need to be easy to use and have a low learning curve for teachers and students \cite{brown2020abstract, bart2018reconciling}.

\section{Instructional strategies (RQ2)}

Table \ref{tab:instructionalstrategies} summarizes the utilized instructional strategies by the reviewed papers and associated supporting tools to apply AI in CT education. 

\begin{table}
    \caption{Instructional strategies and supporting tools used to apply AI in CT education}
\label{tab:instructionalstrategies}
\fontsize{8pt}{8pt}\selectfont
\begin{tabular}{|p{3.0cm}|p{7.0cm}|p{0.7cm}|l|}
\hline
\textbf{Instructional strategy} & \textbf{Supporting tool} & \textbf{Papers} \\
\hline

\multirow{3}{*}{Gamification} & CTQ & \cite{ng2021development} \\
                              \cline{2-3}
                              & GidgetML & \cite{miljanovic2020gidgetml} \\
                              \cline{2-3}
                              & Forest Friends & \cite{sung2017forest} \\
                              \hline

\multirow{4}{*}{\shortstack{Personalized,\\adaptive \\learning}} & A Recommendation system of learning objects & \cite{prisco2017using} \\
                                                    \cline{2-3}
                                                    & A prototype for making students' profile of competencies & \cite{figueiredo2019teaching,figueiredo2019predicting} \\
                                                    \cline{2-3}
                                                    & PRIME based on Blockly & \cite{emerson2020predictive} \\
                                                    \cline{2-3}
                                                    & A Game-Design environment based on Unity & \cite{lalle2021combining} \\
                                                    \hline

\multirow{4}{*}{\shortstack{Project-based \\learning}} & A high-level visual programming system built with Blockly & \cite{brown2020abstract} \\
                                        \cline{2-3}
                                        & AlpacaML and Scratch 3.0 & \cite{zimmermann2020youth} \\
                                        \cline{2-3}
                                        & PDL (Problem Description Language) & \cite{lin2021pdl} \\
                                        \cline{2-3}
                                        & ExposAR – a collaborative cross-device AR system & \cite{lunding2022exposar} \\
                                        \hline

Task-based learning & ScratchThAI: a Scratch tutorial chatbot & \cite{katchapakirin2018architectural} \\
\hline
Experiential learning & An Educational APP & \cite{shih2019integrating} \\
\hline

\end{tabular}
\end{table}

\textit{Gamification} and using digital games for learning purposes was utilized by papers \cite{ng2021development, miljanovic2020gidgetml, sung2017forest} as the instructional strategy to support students' learning of CT. In  \cite{ng2021development} the authors reported on the development of an online adaptive gamification platform and course called Computational Thinking Quest (CTQ) for learning CT. CTQ is created using the Unity game engine and Blockly, a block-based visual programming language, and includes features such as an interactive storyline with animated avatars, mini-games, questions with three levels of difficulty, a leaderboard to motivate active participation, and a course management system. GidgetML, reported in \cite{miljanovic2020gidgetml}, is an adaptive task-based serious game for enhancing first-year programming labs by following sequential instructions to learn new programming and CT concepts and proceeding to the next level. The game consists of 18 levels and is based on the concept of Flow theory, which involves a suitable increase in challenge relative to the player's skill \cite{miljanovic2020gidgetml}. Forest Friends, reported in \cite{sung2017forest}, is a two-player resource strategy video game designed to introduce and learn CT concepts in an engaging and accessible way in informal and public spaces such as museums. The game includes open-ended tasks to engage students in the learning of programming and CT concepts and to practice experimentation, tinkering, and strategic planning skills \cite{sung2017forest}.

\textit{Personalized and adaptive learning} was followed as the instructional strategy to underpin developing tools to support students' learning of CT and programming concepts in a group of the reviewed papers. In \cite{prisco2017using}, a recommendation system of learning objects is created to personalize the teaching and learning of CS concepts as a part of a virtual learning environment. To this end, a student-centered approach combined with the concept of genetic epistemology of Piaget \cite{piaget1972intellectual} was followed to motivate students and catalyze their learning process by allowing them to choose problems, personalize, and improve their learning. The authors of \cite{figueiredo2019teaching} reported on developing a personalized educational tool, based on the concept of video games such as FIFA 19, to construct students' programming competencies profiles and improve their programming skills. Similarly, in \cite{figueiredo2019predicting}, a prototype was created to profile students' programming and CT competencies, such as cognitive reasoning abilities and spatial visualization, and to understand and predict the factors influencing their success or failure in learning these competencies. In \cite{emerson2020predictive} the authors reported on predicting students' models and modeling competencies in PRIME, a Block-Based programming environment based on Google’s Blockly, with Bayesian hierarchical models. Their predictive approach considers individual differences in student characteristics and their programming activities by analyzing the block-based programs created during a series of introductory programming tasks. The study presented in \cite{lalle2021combining} reported on the creation of video games by students in an exploratory game-design environment called Unity-CT to foster their CT skills. Then, the data-driven models of students' video games are combined with Expert (e.g., teacher) knowledge to provide personalized support to foster students' CT Skills.

\textit{Project-based learning} was followed as the overall instructional strategy by studies presented in \cite{brown2020abstract,zimmermann2020youth,lin2021pdl,lunding2022exposar} to apply AI to CT education. In \cite{brown2020abstract} a high-level visual programming system was developed using Blockly and then used by the participants to implement and run their projects (i.e. museum tour) on a real social robot. In the study reported in \cite{zimmermann2020youth}, the participants created and remixed projects to incorporate custom machine-learning models using a toolkit called AlpacaML for generating gesture-controlled interactive media. Further, to scaffold and improve students' problem-solving skills in programming courses, the authors in \cite{lin2021pdl} developed a problem description language (PDL) that decouples problem comprehension and solution development. PDL was evaluated in a project-based programming setting to help students analyze problems, formulate problem descriptions,
and learn about the resulting code generated for solving the formulated problems. In \cite{lunding2022exposar}, the authors developed a collaborative authoring cross-device Augmented Reality (AR) tool to bring Augmented Reality to Computational Thinking through the co-creation of a simple AR application by students.

\textit{Task-based learning} instructional strategy was followed in \cite{katchapakirin2018architectural} to create and introduce ScratchThAI, a Scratch tutorial chatbot. ScratchThAI is meant to assist Thai students with learning and working with Scratch by assigning mission programming tasks and providing hints and feedback, examples, related knowledge, and resources during their learning experience. 

\textit{Experiential learning} theory of Kolb \cite{kolb2014experiential} (i.e., Experiencing, Reflecting, Generalizing, Applying) was combined with the CT concepts
in \cite{shih2019integrating} to develop an education App allowing users to experience and learn artificial intelligence methods anytime, anywhere using mobile devices.

\section{CT components (RQ3)}
As with the definition of CT, the range of proposed CT concepts in the literature is highly diverse. Brennan and Resnick \cite{brennan2012new} defined three categories or dimensions for CT, namely CT concepts, CT practices, and CT perspectives. They then assigned a set of CT components to these dimensions (e.g., sequence, iteration, and condition for CT concepts), with a focus on CT as coding and programming skills. To analyze and identify the covered CT components in the reviewed papers, we took Brennan and Resnick's framework of CT \cite{brennan2012new} as the starting point for our analysis to define an overall structure for the covered CT components by the reviewed papers. Table \ref{tab:CTconcepts} summarizes the identified CT components covered by the reviewed papers.


 
 Compared to Brennan and Resnick's framework for CT \cite{brennan2012new}, our identified list depicts broader coverage and a fresh set of CT concepts, practices, and perspectives, providing a wider understanding of CT beyond just teaching and learning coding and programming skills. For example, we included the notion of \textit{computational empowerment} proposed by Dindler et al. \cite{dindler2020computational} in the CT perspectives. This notion highlights the importance of students' understanding, engagement with, and reflection on different technologies, as well as their ability to make critical and informed decisions about the role of technology in their lives, as part of the CT perspectives.

\begin{table}
    \caption{CT components covered by the reviewed papers}
    \label{tab:CTconcepts}
    \fontsize{8pt}{8pt}\selectfont
\begin{tabular}{|p{0.5in}|p{3.0in}|p{0.8in}|}
\hline
\textbf{Dimension} & \textbf{Components} & \textbf{Papers}\\
\hline
\multirow{7}{*}{Concepts} & Sequence  & \cite{figueiredo2019teaching,figueiredo2019predicting,lalle2021combining}  \\
\cline{2-3}
& Iteration & \cite{ng2021development,katchapakirin2018architectural,figueiredo2019predicting,emerson2020predictive}  \\
\cline{2-3}
& Condition & \cite{ng2021development,sung2017forest,katchapakirin2018architectural,figueiredo2019predicting,emerson2020predictive}  \\
\cline{2-3}
& Events &  \cite{katchapakirin2018architectural}  \\
\cline{2-3}
& Operators and expressions & \cite{ng2021development,emerson2020predictive}  \\
\cline{2-3}
& Data and variables & \cite{ng2021development,katchapakirin2018architectural,emerson2020predictive}  \\
\hline
\multirow{11}{*}{Practices} & Programming, designing or constructing digital products  & \cite{brown2020abstract,zimmermann2020youth,lalle2021combining,lunding2022exposar}  \\
\cline{2-3}

& Pattern recognition & \cite{ng2021development}  \\
\cline{2-3}

& Problem solving, comprehension, decomposition and synthesis  & \cite{ng2021development,prisco2017using,hemberg2019domain,lin2021pdl} \\
\cline{2-3}
& Modeling and simulation & \cite{zimmermann2020youth} \\
\cline{2-3}
& Abstraction and modularizing & \cite{ng2021development,hemberg2019domain,brown2020abstract,emerson2020predictive} \\
\cline{2-3}
& Algorithmic thinking & \cite{ng2021development,figueiredo2019teaching,figueiredo2019predicting,lalle2021combining}  \\
\cline{2-3}

& Data organization, analysis, representation & \cite{bart2018reconciling}  \\
\cline{2-3}
& Testing and debugging & \cite{emerson2020predictive,lalle2021combining}  \\
\cline{2-3}
& Reasoning & \cite{figueiredo2019teaching,figueiredo2019predicting}  \\
\hline
\multirow{3}{*}{Perspectives} & 
 Connecting and collaborating & \cite{lunding2022exposar}  \\
\cline{2-3}
& Computational empowerment & \cite{lunding2022exposar} \\
\hline

\end{tabular}

\end{table}

\section{AI components (RQ4)}
We identified a diverse set of AI components (e.g., methods, techniques, tools) in the reviewed papers, summarized in Table \ref{tab:AItechniques}, as follows:

\begin{table}[]
        \caption{AI components}
    \label{tab:AItechniques}
    \fontsize{8pt}{8pt}\selectfont
    \begin{tabular}{|p{2in}|p{1.25in}|}
     \hline
     \textbf{AI methods and techniques } & \textbf{Papers} \\
     \hline
      Machine Learning   & 
       \cite{emerson2020predictive},
      \cite{lalle2021combining},
      \cite{miljanovic2020gidgetml},
      \cite{ng2021development}  \\
      \hline
      Deep Learning   & 
      \cite{figueiredo2019teaching}, \cite{figueiredo2019predicting},
        \cite{shih2019integrating} \\
      \hline
      Conversational AI  &
      \cite{katchapakirin2018architectural}  \\
      \hline
      Data Science and analytics &
      \cite{bart2018reconciling},
      \cite{prisco2017using} \\
      \hline 
      Human-Computer Interaction  & 
       \cite{han2021exploration},
        \cite{liu2021ai},
      \cite{lunding2022exposar},
      \cite{zimmermann2020youth},
      \cite{sung2017forest}
      \\
      \hline
      Social Robots  & \cite{brown2020abstract} \\
      \hline 

    \end{tabular}

\end{table}
\textbf{Machine Learning}: The study reported in \cite{emerson2020predictive} proposed a predictive student model that uses Bayesian hierarchical modeling taking into consideration both student-level and programming activity-level characteristics. To support personalized learning of CT, the study reported in \cite{lalle2021combining} extended the FUMA framework by designing AI-driven adaptive support in Unity-CT. GidgetML, reported in \cite{miljanovic2020gidgetml}, is and adaptive version of the Gidget programming game and uses ML to modify tasks related to assessing and predicting learners’ programming competencies. 

\textbf{Deep learning}: Figueiredo and Garcia-Penalvo in \cite{figueiredo2019teaching} developed a dynamic learning model for building the profile of individual students and further predicting their failure in acquiring necessary C programming and CT skills. The authors of  \cite{figueiredo2019predicting} continued the work of Figueiredo and Garcia-Penalvo by developing a predictive model to assess the probability of failure based on student profiles using Artificial Neural Networks and Multi Feed-Forward (MFF) network architecture. Shih in \cite{shih2019integrating} explains the development of a mobile application designed to assist users in learning AI methods and application development. The core AI concept in this application is CNN (Convolutional Neural Networks), a deep learning technique based on artificial neural networks that uses mathematical operations like convolutions. Additionally, TensorFlow, one of the most widely used multi-purpose ML frameworks, was employed.

\textbf{Conversational AI}: the authors of \cite{katchapakirin2018architectural} proposed ScratchThAI, a tutorial chatbot developed for learning and working with Scratch, to assist learners through textual conversations and providing relevant advice and learning resources during the learning process. ScratchThAI provides hints, examples, and corresponding resources to enable students to practice the concepts learned while offering personalized mission assignments and experiences.


\textbf{Data Science and analytics}: A critical aspect in the development and deployment of AI in CT education is the availability of relevant educational datasets. These datasets allow the exploration of diverse aspects and finding answers to different activities in various domains using CT components. On this behalf, the study presented in \cite{bart2018reconciling} reports on creating an open-sourced manual for developing pedagogical datasets that is structured as a collection of patterns oriented on collection, analysis, and integration of knowledge that can assist and be used by both novices as well as advance students. Falling into the data science and data analytics category, the study reported in \cite{prisco2017using} presented a recommendation system for learning objects that defines and uses appropriate learning challenges by integrating learning models into game models. The proposed solution, built upon Online Judges (online platforms supporting programming events and competitions), utilized data analytics to tailor students' learning to their needs and interests. Furthermore, the system automatically identifies components that enhance users’ learning and motivation while providing automated feedback.

\textbf{Human-Computer Interaction}: study presented in \cite{lunding2022exposar} reported on developing ExposAR, a collaborative cross-device AR system that allows learners to create and use AR applications and further reflect on AR technologies.  ExposAR implicitly incorporates AI technologies into various architectural components, focusing on the design of computational concepts, collaboration, and embedding elements from the user’s world into the application. The study presented in \cite{zimmermann2020youth} explained AlpacaML, a machine learning (ML) modeling toolkit with a programming application for building, testing, and evaluating ML models of gestures based on wearable sensor data. This application was further connected with Scratch for building and executing modeling tasks using Scratch blocks. During this process, users choose movements to be modeled, place the sensors in the correct positions, collect data using a phone's camera, and build the model where users define the training set and label the data. They then continue by building the model for classifying time series. The model is executed and evaluated, allowing users to assess its performance and make necessary adjustments. The study presented in \cite{sung2017forest} introduced the Forest Friends strategic video game to teach AI coding and CT components in an accessible and engaging manner. In this solution, users learn computational skills and gain experience with conditional statements. 




\textbf{Social robots}: Brown et al. in \cite{brown2020abstract} presented a high-level visual programming system for designing robot tasks by novices by incorporating social behavioral cues and being deployed in a museum. The proposed system is developed in Blocky, and the robot acts as a tour guide in the museum. Users can design a museum tour that considers both high-level technical aspects and social behavioral aspects.

\section{Discussion and conclusions}


%

Through this systematic literature review, we explored various aspects of utilizing AI to support Computational Thinking (CT) education in higher education, namely, its benefits and challenges, employed instructional strategies, covered CT components, and applied AI techniques and methods. 

The study revealed various benefits and challenges of utilizing AI in CT education. The identified benefits include the high potential of AI to reshape CT education by personalizing learning and addressing students' diverse and heterogeneous CT and programming backgrounds, which is a known challenge in CT and programming education (e.g., see \cite{lahtinen2005study}). On the other hand, the identified challenges include the complexity of characterizing and modeling students' CT backgrounds and competencies needed by AI-based predictive models, the AI-related knowledge and expertise requirements for teachers, and the AI's potential to diminish students' creativity and innovation due to their over-reliance on AI. While recognizing the importance of the first two aforementioned challenges, we conclude that, given the nature of CT as a set of thinking skills and the rapid advancements in the AI domain, addressing the latter challenge (i.e., AI's potential to diminish students' creativity and innovation) requires thorough and emergent research. Such research should investigate the extent and mechanisms that AI may shape or replace students' creative and computational thinking, as well as its pedagogical implications for CT education, a call also made by other researchers (e.g., \cite{han2021exploration} and  \cite{tedre2021ct}).

A notable finding from our study concerns the identified CT components (see Table \ref{tab:CTconcepts}). While the most frequently covered CT concepts were programming-related topics such as condition, iteration, and sequence, we identified a fresh set of CT practices, including designing and constructing digital products, data organization and analysis, and reasoning, alongside established CT practices such as problem comprehension, decomposition, problem-solving, algorithmic thinking, and abstraction. Furthermore, we expanded the existing CT perspectives dimension by identifying and incorporating the notion of computational empowerment, which emphasizes that CT should not only focus on teaching basic programming skills but also enable students to understand, engage with, and reflect on emerging digital technologies, such as AI and augmented reality \cite{lunding2022exposar,dindler2020computational}. This notion appears to have a strong connection to the recent \textit{decoding approach} in CT education, which aims to teach and enable students to learn how to 'decode' computer and digital models, systems, and tools (e.g., AI-based tools) to investigate the processes embodied in their code \cite{lee2024decoding} and reason about their outputs. Our conclusion here is that, while CT components have so far been dominated by programming- and coding-related concepts, practices, and skills, the emergence of new digital technologies and their growing role in everyday life necessitate recognizing new types of CT concepts and practices that focus on developing students' 'decoding' skills.

Among the utilized AI techniques and methods, as presented in Table \ref{tab:AItechniques}, are various supervised and unsupervised ML techniques, such as Bayesian modeling and k-means, as well as social robots, AR technologies, and conversational AI to facilitate educational activities and initiatives to build students' CT skills. Moreover, some data science methods, such as defining and using educational datasets and data analytics, were utilized.

\textbf{Limitations}:
This systematic literature review examined academic publications on AI applications for CT education in higher education, published between 2017 and 2022 in the ACM and IEEE digital libraries. We deliberately selected this time frame to avoid overlap with the emergence of publications on Generative AI and LLMs from 2023. We recommend further research built upon this study by expanding the time frame and including additional outlets such as Springer, Elsevier, Scopus, and Web of Science.





\bibliographystyle{acm}

\bibliography{references}

\end{document}